\documentclass[fleqn,usenatbib]{mnras}
\usepackage{graphics}
\usepackage{graphicx}
\usepackage{color}
\usepackage{aas_macros}
\usepackage{natbib}
\usepackage{amsmath}
\usepackage{comment}
\usepackage{soul} 

\newcommand{\unit}[1]{\textnormal{#1}}
\newcommand{\Hunit}{\ensuremath{\frac{\unit{km}/\unit{s}}{\unit{Mpc}}}}

\title[The ISW effect in the AvERA cosmology]{The integrated Sachs--Wolfe effect in the AvERA cosmology}

\author[R. Beck et al.]{R\'obert Beck$^{1,2}$\thanks{E-mail: beckrob@ifa.hawaii.edu}, Istv\'an Csabai$^2$, G\'abor R\'acz$^2$, Istv\'an Szapudi$^1$
\\
$^1$Institute for Astronomy, University of Hawaii, 2680 Woodlawn Drive, Honolulu, HI, 96822 \\
$^2$Department of Physics of Complex Systems, E\"{o}tv\"{o}s Lor\'and University, Pf. 32, H-1518 Budapest, Hungary
}

\date{Accepted XXX. Received YYY; in original form ZZZ}

\pubyear{2018}

\begin{document}
\label{firstpage}
\pagerange{\pageref{firstpage}--\pageref{lastpage}}
\maketitle

\begin{abstract}

The recent AvERA cosmological simulation of \citet{Racz2017} has a  $\Lambda \mathrm{CDM}$-like  expansion history and removes the tension between local and Planck (cosmic microwave background) Hubble constants. We contrast the AvERA prediction of the integrated Sachs--Wolfe (ISW) effect with that of  $\Lambda \mathrm{CDM}$. The linear ISW effect is proportional to the derivative of the growth function, thus it is sensitive to small differences in the expansion histories of the respective models.
We create simulated ISW maps tracing the path of light-rays through the Millennium XXL cosmological simulation, and perform theoretical calculations of the ISW power spectrum.
AvERA predicts a significantly higher ISW effect than $\Lambda \mathrm{CDM}$, $A=1.93-5.29$ times larger depending on the $l$ index of the spherical power spectrum, which could be utilized to definitively differentiate the models. We also show that AvERA predicts an opposite-sign ISW effect in the redshift range $z \approx 1.5 - 4.4$, in clear contrast with  $\Lambda \mathrm{CDM}$. Finally, we compare our ISW predictions with previous observations. While at present these cannot distinguish between the two models due to large error bars, and lack of internal consistency suggesting systematics, ISW probes from future surveys will tightly constrain the models.

\end{abstract}

\begin{keywords}
cosmology: large-scale structure of Universe -- cosmology: dark energy -- cosmological parameters -- methods: numerical.
\end{keywords}

\maketitle

\section{Introduction}

The ,,standard model of cosmology", the Lambda cold dark matter ($\Lambda \mathrm{CDM}$) model enjoys enormous success in explaining a wide variety of observations over a range of scales. However, in the past few years an intriguing point of tension emerged between expectations from the standard $\Lambda \mathrm{CDM}$ model and measurements of the local value of the Hubble constant. \citet{Riess2011} and \citet{Riess2016} calibrated the cosmic distance ladder from geometric distances through Cepheids to supernovae, yielding the most recent value of $H_0=73.24\pm1.74~\Hunit$. In contrast to this is the $H_0=67.51\pm0.64~\Hunit$ parameter value of the $\Lambda \mathrm{CDM}$ model that achieved the best fit to the cosmic microwave background (CMB) as measured by the Planck satellite \citep{Planck2015}. While the disparity is not entirely definitive, it is over $3\sigma$ and is thus worthy of scrutiny.

Recently, the AvERA inhomogeneous cosmological simulation of \citet{Racz2017} claimed to naturally reconcile local and CMB measurements of the Hubble constant. It yielded a very similar expansion history to that of $\Lambda \mathrm{CDM}$, consistent with most observational constraints, without assuming dark energy. Naturally, the question arises whether --- and if so, how --- the AvERA and $\Lambda \mathrm{CDM}$ cosmologies can be distinguished by observational probes.

A prime candidate for such a cosmological probe is the integrated Sachs--Wolfe effect, which establishes a link between observations of large-scale structure and of photons comprising the CMB. As photons propagate in an expanding universe, they undergo a small energy change as they pass through the decaying potential wells and peaks of galaxy clusters and voids, respectively, leading to a small, secondary anisotropy in the CMB temperature map.
The physical description of this phenomenon is given by the integrated Sachs--Wolfe and Rees--Sciama effects \citep[caused by the linear and non-linear growth of structure, respectively;][]{Sachs1967,Rees1968}.
Since these effects are sensitive to the derivative of the growth function, small differences in expansion history may be amplified to yield significantly differing expected signals. In fact, the dark-energy-free DGP cosmological model of \mbox{\citet{Dvali2000}} has been shown to be statistically unlikely due to its ISW signature \citep{Fang2008}.

While small in magnitude, observational evidence from the past decade has demonstrated the significance of the late-time integrated Sachs--Wolfe effect (hereafter referred to as ISW).
There have been two main approaches in the literature, either via the stacking of voids and clusters \citep{Granett2008,Papai2010,Papai2011,Granett2015,Planck2016,Nadathur2016,Kovacs2017,Cai2017}, or by cross-correlating galaxy surveys with the CMB map \citep{Peiris2000,Scranton2003,Ho2008,Giannantonio2008,Raccanelli2008,Granett2009,Francis2010,Goto2012,Giannantonio2012,Granett2015,Planck2016}. 
Both approaches eventually led to statistically significant detections of the effect, with the highest significances over $4$ sigmas \citep{Granett2008,Giannantonio2008,Giannantonio2012}. 

Most of the measurements quoted above are statistically consistent with expectations from the standard $\Lambda \mathrm{CDM}$ model, however, it cannot be stated that observational consensus has been reached in the field. Several experiments detected an ISW signal amplitude that exceeds the $\Lambda \mathrm{CDM}$ prediction by $1$ sigma \citep{Giannantonio2008,Granett2009,Giannantonio2012,Granett2015,Nadathur2016}, $2$ sigmas \citep{Granett2008,Ho2008,Papai2011,Goto2012,Kovacs2017} or even $3$ sigmas \citep{Cai2017}, even though some of the detections were later revised to be of smaller magnitude as new data came available \citep{Kovacs2013,Ferraro2015,Shajib2016}. The reported $A$ amplitude factors ($A=1$ matches $\Lambda \mathrm{CDM}$) also vary wildly among papers, from just $0.64$ up to even $20$, depending on the dataset and methodology details. Further research is needed to resolve these inconsistencies.

Despite the present observational uncertainties, as the maximum achievable ISW detection significance was estimated to be over $7 \sigma$ \citep{Afshordi2004}, there is value in investigating the predictions of cosmological models alternative to $\Lambda \mathrm{CDM}$ with respect to the elusive ISW effect. The detection of the effect has usually been quoted as proof of the existence of dark energy, but that statement is only true within the framework of standard homogeneous $\Lambda \mathrm{CDM}$ cosmologies --- more precisely, the existence of the ISW signal depends on the properties of the growth function. The AvERA algorithm of \citet{Racz2017} assumed no dark energy, and while it resolved the Hubble constant tension, it has not yet been shown whether it can give reasonable predictions for the ISW effect.

Thus, the main goal of the present paper is to investigate this issue, and compare the magnitude of the ISW effect predicted by the standard $\Lambda \mathrm{CDM}$ cosmological model and by the AvERA cosmological 
simulation. To achieve this, we trace light-rays through the Millennium XXL simulation of \citet{Angulo2012}, yielding a direct evaluation of the ISW effect, and we also compute the theoretical expectation from the two models using their respective linear growth functions. Finally, we contrast the predictions with previous measurements in the literature.

\section{Theory of the integrated Sachs--Wolfe effect}

\label{sec:theory}

The relationship between the derivative of the $\Phi$ gravitational potential and the corresponding temperature shift $\Delta T_{\mathrm{ISW}}$ of the CMB can be expressed with the integral
\begin{equation}
\frac{\Delta T_{\mathrm{ISW}}}{T_{\mathrm{CMB}}} = - \frac{2}{c^2} \int_{\tau_0}^{\tau_{\mathrm{CMB}}} d\tau \frac{d\Phi\left(\vec{x}(\tau),\tau\right)}{d\tau}
\label{eq:ISWintegral}
\end{equation}
over the path of a CMB photon, from present conformal time $\tau=\tau_0$ to the surface of last scattering where $\tau=\tau_{\mathrm{CMB}}$ \citep[e.g.][]{Sachs1967,Seljak1996,Cai2009,Cai2010}. $T_{\mathrm{CMB}}$ is the mean temperature of the CMB, $c$ is the speed of light, and $\vec{x}$ denotes comoving coordinates. 

For the gravitational potential, we have the Poisson equation 
\begin{equation}
\vec{\nabla}^2\Phi\left(\vec{x},\tau\right)=4\pi G \overline{\rho}(\tau)a^2(\tau) \delta \left(\vec{x},\tau \right),
\end{equation} 
where $G$ is the gravitational constant, $\overline{\rho}$ is the average density of the universe, $a$ is the scale factor and $\delta=(\rho-\overline{\rho}) / \overline{\rho}$ is the density contrast. In Fourier space, this can be written as 
\begin{equation}
\Phi\left(\vec{k},\tau\right)=-\frac{3}{2} H_0^2 \Omega_m \frac{1}{a(\tau)} \frac{\delta\left(\vec{k},\tau\right)}{k^2},
\end{equation}
where we substituted $4 \pi G \overline{\rho} = \frac{3}{2} H_0^2 \Omega_m$, introducing the $H_0$ Hubble constant and the present value of the matter density parameter, $\Omega_m$. Using the continuity equation in Fourier space, $\dot{\delta} + i\vec{k}\vec{p}=0$, with overdot representing derivative with respect to cosmic time, the derivative of the potential can be expressed as
\begin{equation}
\frac{d\Phi\left(\vec{k}(\tau),\tau\right)}{d\tau} = \frac{3}{2} \frac{H_0^2}{k^2}\Omega_m \left[ H(\tau)\delta\left(\vec{k},\tau\right) + i \vec{k}\vec{p}\left(\vec{k},\tau \right) \right],
\label{eq:potderiv}
\end{equation}
where $\vec{p}\left(\vec{k},\tau\right)$ is the Fourier transform of $\vec{p}\left(\vec{x},\tau\right)=\left(1+\delta(\vec{x},\tau)\right)\vec{v}\left(\vec{x},\tau \right)$ --- the momentum density divided by the mean mass density ---, and $H(\tau)=\dot{a}/a$ is the Hubble parameter. 

The full expression Eq.~\ref{eq:potderiv} of the derivative of the potential cannot be computed when momentum information is not available. In such cases, the assumption of linear growth can be adopted, $\delta\left(\vec{k},\tau\right)=D(\tau)\delta\left(\vec{k},\tau=\tau_0\right)$, using the linear growth factor $D(\tau)$. Replacing the continuity equation form of the derivative of $\delta$ with this approximation we get
\begin{equation}
\frac{d\Phi\left(\vec{k}(\tau),\tau\right)}{d\tau} = \frac{3}{2} \frac{H_0^2}{k^2}\Omega_m \left[ H(\tau)\delta\left(\vec{k},\tau\right) \left(1-\beta(\tau)\right) \right],
\label{eq:potderivlin}
\end{equation}
where we introduced $\beta(\tau) \equiv d \ln D / d \ln a $. Note that $\beta(\tau)=1$ in a zero dark energy homogeneous $\mathrm{CDM}$ cosmological model, therefore there is no ISW effect therein. However, in any cosmology where generally $\beta(\tau) \neq 1$, there will be an ISW effect --- such as in AvERA, where there is also no dark energy, but the unusual growth factor behavior is instead caused by the way the inhomogeneities are taken into account (see Sect.~\ref{sec:cosmo}).

Thus, from a known density field, and in a given cosmological model (Hubble parameter and growth factor), the ISW effect may be computed by tracing the path of light through that field using Eqs.~\ref{eq:ISWintegral} and \ref{eq:potderivlin}. The resulting ISW map may then be transformed via harmonic analysis into a $C_l^{\mathrm{ISW}}$ spherical autocorrelation power spectrum.

In linear theory, an alternative way to arrive at the ISW auto power spectrum from strictly theoretical considerations requires the computation of the matter overdensity power spectrum at present time, $P_\delta (k)$. There are fast and efficient public codes that perform this task in a Boltzmann equation framework. e.g. CMBFAST \citep{Seljak1996b} and CAMB \citep{Lewis2000,Lewis2002,Challinor2005}. The $C_l$ power spectrum elements can be expressed by

\begin{equation}
C_l^{\mathrm{ISW}} = \frac{2}{\pi} \int dk \, k^2 P_\delta (k) \left[ G_l^{\mathrm{ISW}}(k) \right] \left[ G_l^{\mathrm{ISW}}(k) \right],
\label{eq:sphericalpower}
\end{equation}

where $G_l^{\mathrm{ISW}}$ is the following expression associated with the ISW effect \citep[e.g.]{Peiris2000,Ho2008},

\begin{multline}
\left[ G_l^{\mathrm{ISW}}(k) \right] = \\ = T_{\mathrm{CMB}} \left(\frac{3 H_0^2\Omega_m}{c^2}\right)\int d \tau \frac{d}{d\tau}\left( \frac{D}{a} \right) j_l [\chi(\tau) k] \frac{1}{k^2}.
\label{eq:bessel}
\end{multline}

Above, $\chi(\tau)=c(\tau_0-\tau)$ is the conformal lookback distance, and $j_l$ denotes the spherical Bessel function of the first kind.
We note here that while many works utilize the small angle or Limber approximation \citep{Limber1953,Kaiser1992} to speed up the computation of Eq.~\ref{eq:bessel}, we instead calculate the full formula. Refer to Appendix~\ref{sec:appendix1} for more details regarding the accuracy of the Limber approximation in the context of AvERA cosmology.

\section{The AvERA cosmological simulation}

\label{sec:cosmo}

The AvERA (Average Expansion Rate Approximation) cosmological simulation of \citep{Racz2017} takes into account local inhomogeneities using the separate universe conjecture, and performs volume averaging inspired by the Buchert equations \citep{Buchert2000,Buchert2001,Buchert2012} to yield an \textit{effectively} homogeneous expansion model.

The key point of the algorithm that separates it from standard approaches is that the order of volume-averaging and expansion rate computation is switched up, as illustrated in Fig.~\ref{fig:scheme1}. In summary, small local regions expand according to their local density and the Friedmann equations, and the corresponding scale factor increment of the universe is a volume average of the local scale factor increments. Additionally, in AvERA no dark energy was assumed, i.e. $\Omega_\Lambda = 0$.

\begin{figure}
\begin{center}

\begin{equation}
\sbox0{$\begin{array}{ll}
        \Omega_{m,1}\\
        \Omega_{m,2}\\
        \dots \\
        \Omega_{m,N}\\
\end{array}$}
\mathopen{\resizebox{1.2\width}{\ht0}{$\Bigg\langle$}}
\usebox{0}
\mathclose{\resizebox{1.2\width}{\ht0}{$\Bigg\rangle$}} 
\Rightarrow \boxed{\mathrm {Friedmann~eq.}} \Rightarrow
\sbox1{$\begin{array}{ll}
                  \Delta V \\
\end{array}$}
\usebox{1}
\Rightarrow a(t+\delta t)
\label{eq:averaging1}
\end{equation}

\begin{equation}
\sbox0{$\begin{array}{ll}
                  \Omega_{m,1}\\
 		\Omega_{m,2}\\
		\dots \\
		 \Omega_{m,N}\\
\end{array}$}
\usebox{0}
\Rightarrow \boxed{\mathrm {Friedmann~eq.}} \Rightarrow
\sbox1{$\begin{array}{ll}
                 \Delta V_1\\
 		\Delta V_2\\
		\dots \\
		\Delta V_N\\
\end{array}$}
\mathopen{\resizebox{1.2\width}{\ht0}{$\Bigg\langle$}}
\usebox{1}
\mathclose{\resizebox{1.2\width}{\ht0}{$\Bigg\rangle$}} 
\Rightarrow a(t+\delta t)
\label{eq:averaging2}
\end{equation}

\end{center}
\caption{Top: Standard cosmological N-body simulations evolve the Friedmann equations using the average density. Since the total mass is constant the scale factor increment is independent of density fluctuations. Bottom: The AvERA simulation calculates the expansion rate of local mini-universes and averages the volume increment spatially to get the global scale factor increment.}
\label{fig:scheme1}
\end{figure}

Results from AvERA showed a similar expansion history to the standard $\Lambda \mathrm{CDM}$ model, but at the same time the relationship between the Hubble parameter at the epoch of recombination and at present time was different from that of $\Lambda \mathrm{CDM}$. 
This relationship was also sensitive to the particle mass, which is a ,,free'' input parameter of AvERA, and can be thought of as the mass of gravitationally bound mini-universes.
Fitting the initial conditions of the simulation to the Planck best-fit $\Lambda \mathrm{CDM}$ cosmology \citep{Planck2015}, a reasonable particle mass of $1.17 \times 10^{11} M_\odot$ 
yielded a present-time Hubble parameter of $H_0=73.1 \Hunit$, matching the local measurement of \citep{Riess2016}, thus resolving the tension between local and CMB-based measurements of the Hubble parameter. See \citet{Racz2017} for more details.

The AvERA simulation had a linear simulation box size of $147.623 \textrm{Mpc}$, and assumed periodic boundary conditions. This very limited volume is not sufficient for ISW calculations using Eq.~\ref{eq:ISWintegral},
as the periodic boundary conditions cause many repetitions of the same structures along the path of light-rays in certain directions, leading to strong artifacts. Instead, we only extracted the effective scale factor --- and functions derived therefrom --- from AvERA, and performed raytracing in a standard N-body simulation of much larger volume, Millennium XXL \citep{Angulo2012}.

The cosmological information needed for the ISW computation (see Eqs.~\ref{eq:potderivlin}--\ref{eq:bessel}) is contained in the functions $\beta(\tau) = d \ln D / d \ln a $, the related term $\frac{d}{d\tau}\left( \frac{D}{a} \right)$, and $H(\tau)$.

The $t$ time, $H$ Hubble parameter and effective $\Omega_m$ outputs of the simulation were interpolated as functions of the $a$ scale factor using a cubic spline. The scale factor was sampled between the initial value of $0.1047$ and the present-time value of $1.0$ with a linear step size of $0.001$. From the interpolated functions, the $\tau$ conformal time, $D$ growth factor, and the derivatives $\beta$ and $\frac{d}{d\tau}\left( \frac{D}{a} \right)$ were calculated numerically.

In computing the initial value of the conformal time, we assumed a simple Einstein--de Sitter cosmology, as in such early times the cosmological models do not differ noticeably.
Since the numerical derivatives of the (functions of the) discretely sampled scale factor were relatively noisy, we performed Gaussian smoothing with a $\sigma$ of $10$ data points, i.e. $\sigma_a=0.01$.

\begin{figure*}
\begin{center}
\includegraphics[width=\textwidth]{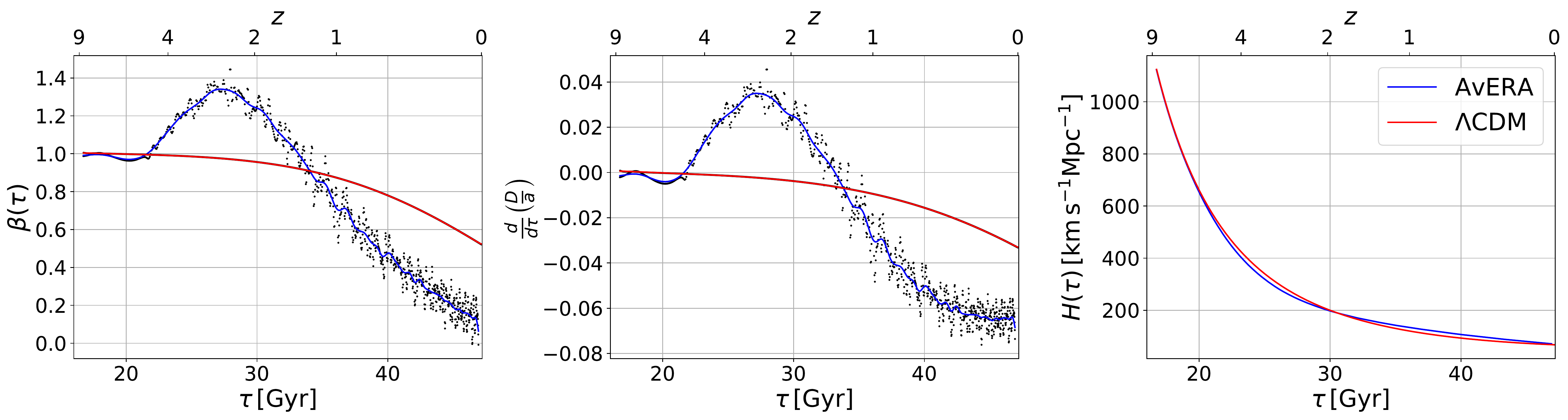}
\vspace*{-0.5cm}
\end{center}
\caption{The cosmological functions utilized in the ISW calculations, extracted from the AvERA inhomogeneous simulation \citep[blue,][]{Racz2017}, and the corresponding values for a standard $\Lambda \mathrm{CDM}$ model with the best-fit parameters from Planck \citep[red,][]{Planck2015}. Black points show the raw numerical derivatives, while the continuous curves have Gaussian smoothing applied.}
\label{fig:cosmo}
\end{figure*}

Fig.~\ref{fig:cosmo} shows the resulting relevant cosmological functions for the AvERA algorithm and the $\Lambda \mathrm{CDM}$ model. Of note are the larger values of the derivatives in the case of the AvERA algorithm, foreshadowing a larger amplitude for the predicted ISW effect.
Additionally, an interesting feature of the AvERA cosmology is the positive growth factor derivative between $\tau \approx 22-33 \, \mathrm{Gyr}$, corresponding to the redshift range $z \approx 1.5 - 4.4$. This would imply an opposite-sign ISW effect within this redshift range, and it also affects the accuracy of the Limber approximation (see Appendix~\ref{sec:appendix1}). In fact, this effect could be used as a \textit{smoking gun} to test the validity of the AvERA model --- the sign of the matter--CMB correlation at this specific redshift range, obtained e.g. by using a quasar sample, could clearly prove a departure from $\Lambda \mathrm{CDM}$, or alternatively disprove the prediction of AvERA.

\section{Raytracing in Millennium XXL}

\subsection{Motivation}
 
As described in Sect.~\ref{sec:theory}, within linear theory, as long as only the ISW power spectrum is required, there is no need to perform a computationally intensive raytracing procedure, instead the formulas in Eqs.~\ref{eq:sphericalpower}--\ref{eq:bessel} can be directly evaluated --- indeed, we have done so, see Sect.~\ref{sec:amplitude} for details. Additionally, as we will see in Sect.~\ref{sec:tracedetails}, we have to adopt the assumption of linear growth in our calculations.

Still, there are distinct advantages to computing the entire temperature map from an actual density realization via raytracing. First, a map allows a visual inspection of features. Second, it contains more information, and thus enables the computation of more complicated measures, e.g. the three-point correlation function, or a simulated stacking analysis. Third, and most importantly, the procedure allows a semi-independent cross-verification of theoretical results by directly computing the power spectrum from the map. Matching results from the two different approaches would stoke confidence in their validity.

\subsection{The Millennium XXL simulation}

The Millennium XXL simulation of \citet{Angulo2012} is a larger-scale version of the earlier Millennium simulation \citep{Springel2005,Springel2005b}. It tracked the movement of $303,464,448,000$ dark matter particles of mass $8.456 \times 10^9 \, M_\odot$ in a comoving volume of $\left(4.1096 \, \mathrm{Gpc}\right)^3$. This large volume is suitable for ISW analyses, since the light-ray reaches the boundary of the simulation box for the first time at a redshift of $z \approx 1.3$, thus the assumed periodic boundary conditions do not affect results significantly.

Particle density fields, computed from particle counts in cells, are available in a grid of $1024^3$ points (corresponding to a linear cell size of $4.013 \, \mathrm{Mpc}$) for $51$ snapshots between $z=9.28$ and $z=0$. Velocity information is not available in these snapshots.

\subsection{Raytracing details}

\label{sec:tracedetails}

Since the Millennium XXL is a $\Lambda \mathrm{CDM}$ simulation, with a slightly different expansion history from that of AvERA, the simulated density field cannot be paired to AvERA cosmological parameters along its evolution. However, a reasonable solution is to use an early snapshot of the simulation, when inhomogeneities are still small and different cosmologies match closely, and apply the linear growth approximation to evolve the density field from this early time following the growth function. In any case, as only density data is available for the snapshots, the linear growth approximation cannot be avoided. Thus, we use the first Millennium XXL snapshot within the redshift coverage of the AvERA run, at $z=8.55$, and perform raytracing in the corresponding simulation box to evaluate the ISW effect, for both the standard $\Lambda \mathrm{CDM}$ and AvERA cosmologies.

Adopting the assumption of linear growth, we calculated Eq.~\ref{eq:ISWintegral} via Eq.~\ref{eq:potderivlin}. Light-rays were projected from $3$ randomly selected $z=0$ starting positions ,,into the past'' in $12 \times 64^2$ directions, matching the HEALPix\footnote{http://healpix.sourceforge.net/} \citep{Gorski2005} spherical coordinates with $\mathrm{NSIDE}=64$, up to $z=8.55$. The derivative of the gravitational potential was evaluated at locations spaced $0.75 \, \mathrm{Mpc}$ apart (in comoving coordinates) by performing trilinear interpolation within grid points. Then, these local contributions were numerically integrated along each light-curve to yield an ISW temperature map for every starting position and cosmology.

\section{Results}

\subsection{ISW amplitude}

\label{sec:amplitude}

\begin{figure*}
\begin{center}
\includegraphics[width=0.8\textwidth]{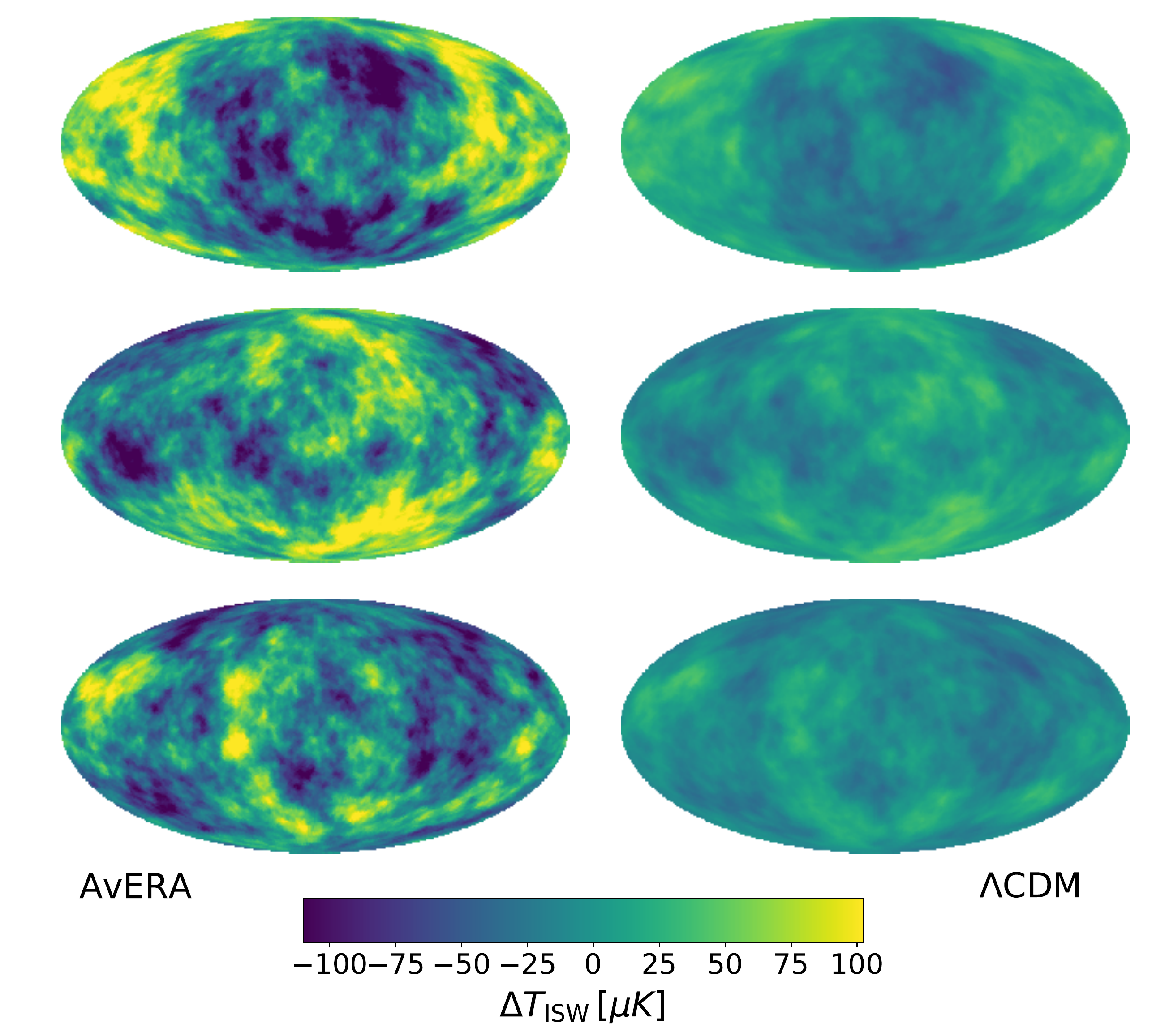}
\vspace*{-0.5cm}
\end{center}
\caption{The ISW temperature maps corresponding to the AvERA (left) and the standard $\Lambda \mathrm{CDM}$ cosmology (right), integrated from $z=0$ to $z=8.55$ by raytracing through the Millennium XXL simulation using the linear growth approximation. Three different, randomly selected starting locations for the raytracing --- the same three for the two cosmologies --- are presented to illustrate cosmic variance.}
\label{fig:ISWmap}
\end{figure*}

The $\Delta T_{\mathrm{ISW}}$ (see Eq.~\ref{eq:ISWintegral}) maps resulting from the raytracing analysis are presented in Fig.~\ref{fig:ISWmap}. While the same overall structures are visible in the case of both the standard $\Lambda \mathrm{CDM}$ and AvERA cosmologies, the amplitude of the ISW effect is clearly much larger for AvERA.

To contrast the ISW amplitudes of the two cosmologies in a detailed way, we computed the ISW autocorrelation spherical power spectra from the maps using the Anafast function of Healpy/HEALPix\footnote{https://github.com/healpy/healpy} \citep{Gorski2005}.

Additionally, we evaluated Eqs.~\ref{eq:sphericalpower}-\ref{eq:bessel} directly, yielding a theoretical expectation to independently verify the raytracing results.
The $P_\delta (k)$ linear matter power spectrum was obtained from PyCAMB\footnote{http://camb.readthedocs.io/en/latest/} \citep{Lewis2002} using the following cosmological parameters \citep[matching those of][see Table 4 in that paper for more information]{Planck2015}: $H_0=67.74$, $\Omega_b h^2=0.0223$, $\Omega_c h^2=0.1188$, $\tau = 0.066$, $n_s=0.9667$, spatially flat geometry and no contribution from tensor modes, i.e. $\Omega_k=0$ and $r_{0.002}=0$. Additionally, a single massive neutrino of mass $m_{\nu}=0.06 \, \mathrm{eV}$ was assumed.

Similarly to the raytracing case, the matter power was evaluated at $z=8.55$ and then scaled with the growth factor of the appropriate cosmology. We note here that the Millennium XXL initial matter power spectrum had been derived from different parameters, and to account for this we scaled the raytracing results with the median of the ratio between the standard $\Lambda \mathrm{CDM}$ and the empirical Millennium XXL initial matter power spectrum, $P_{\delta,\Lambda \mathrm{CDM}}/P_{\delta,\mathrm{Millennium}}=0.803$.

With the $P_\delta (k)$ power spectrum and the growth factor derivative $\frac{d}{d\tau}\left( \frac{D}{a} \right)$ known, we numerically integrated Eqs.~\ref{eq:sphericalpower}-\ref{eq:bessel} for the AvERA and Planck $\Lambda \mathrm{CDM}$ cosmologies. The resolution in $\tau$ followed that of the cosmologies, as described in Sect.~\ref{sec:cosmo}, while $k$ was evaluated between $7.05 \times 10^{-6} \, \mathrm{Mpc}^{-1}$ and $1.64 \times 10^{3} \, \mathrm{Mpc}^{-1}$ in $1200$ logarithmically spaced points, covering the relevant range of the power spectrum. Doubling the resolution in both integrals did not change results noticeably, but significantly increased computation time, therefore the current resolution is sufficient for our purposes.

\begin{figure}
\begin{center}
\includegraphics[width=\columnwidth]{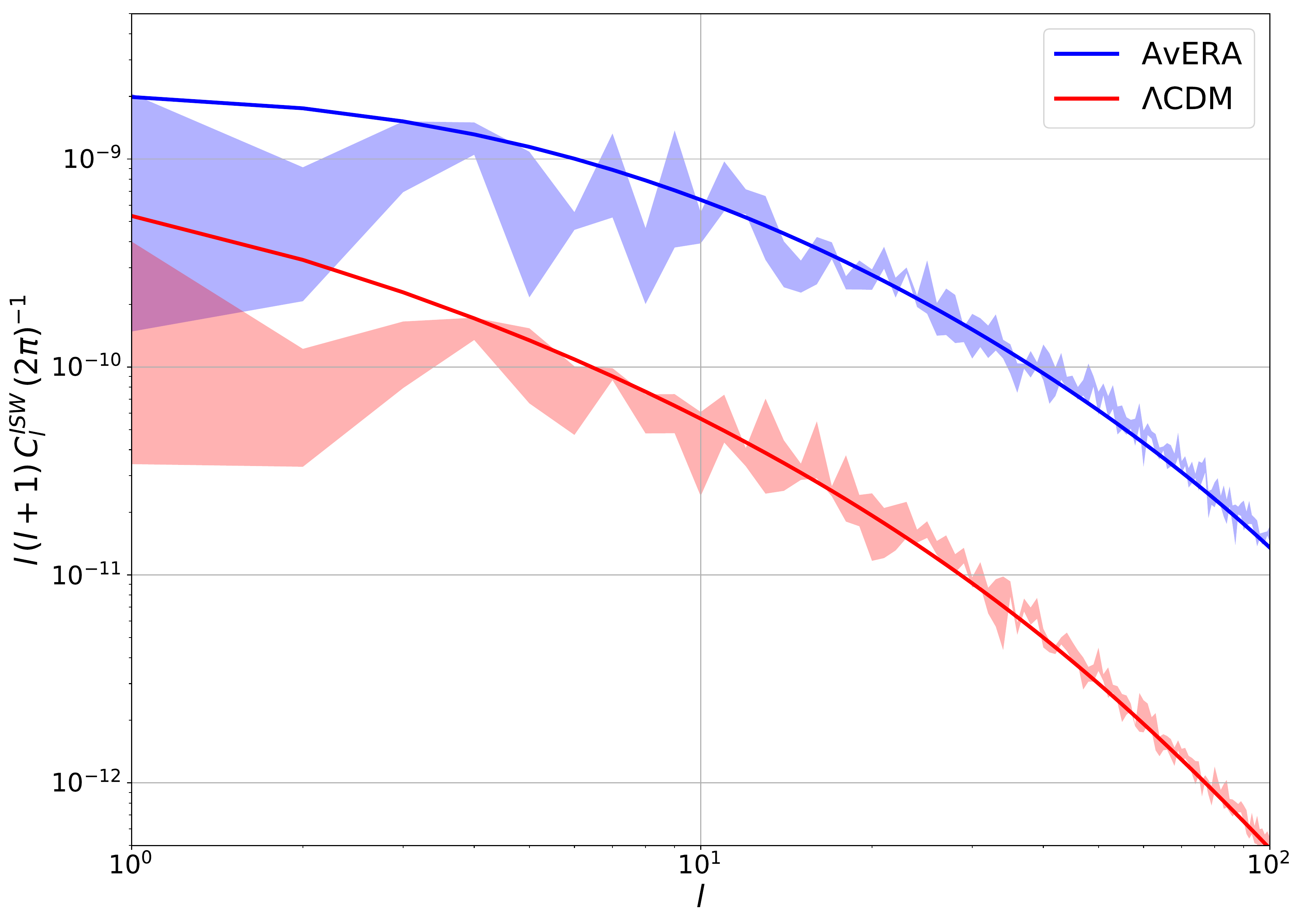}
\vspace*{-0.5cm}
\end{center}
\caption{ISW autocorrelation spherical power spectra for the AvERA (blue) and the standard $\Lambda \mathrm{CDM}$ cosmology (red). The shaded areas correspond to the minimum--maximum range of the spectra at the three raytracing starting locations, while the continuous curves show the results of the theoretical calculation.}
\label{fig:ISWpower}
\end{figure}

The $C_l^{\mathrm{ISW}}$ raytracing and theoretical ISW autocorrelation spherical power spectra are presented in Fig.~\ref{fig:ISWpower}. We only show the range $l=1-100$ as the linear growth approximation (which was applied in both analyses) starts to become inaccurate around $l=70$ \citep{Cai2009,Cai2010}.
We found that the results of the two approaches matched rather well, except at very large scales (or low $l$ values), where the effect of cosmic variance grows significantly, and the finite size of the simulation box starts to become an issue, therefore mismatch is expected in that region. The theoretical and raytracing methods thus support the validity of each other, in the case of both cosmologies. The consistent results also mean that, at least in the context of linear theory, the theoretical calculation is sufficiently accurate to describe the phenomenon.

Hereafter we use the more stable theoretical integration results to discuss the relationships between the power spectra. The ratio of power between the AvERA and $\Lambda \mathrm{CDM}$ curves goes from $3.73$ at $l=1$ to $28.0$ at $l=100$, which corresponds to an $A$ amplitude factor (relative to $\Lambda \mathrm{CDM}$) range of $A=1.93-5.29$. Thus, we can conclude that the AvERA algorithm predicts a significantly larger ISW effect than the standard cosmological model, and that this relationship is strongly dependent on $l$.

\subsection{Comparison with earlier results}

\label{sec:literature}

Numerous works in the literature focused on the observational measurement of the ISW effect, and naturally the ultimate goal would be to observationally determine whether the higher prediction of the AvERA cosmology is tenable, and if yes, which cosmological model fits the data better. However, as we shall soon see, the relatively high error bars of measurements, and the lack of consensus between different approaches prevents a quantitative meta-analysis.

We compiled a list of a number of previous measurements of the magnitude of the ISW effect. The cross-correlation measurements we considered include \citet{Giannantonio2008,Ho2008,Granett2009,Giannantonio2012,Goto2012,Kovacs2013,Ferraro2015,Granett2015,Planck2016,Shajib2016}, while the stacking measurements are comprised of \citet{Granett2008,Papai2011,Granett2015,Cai2017,Kovacs2017}. Of these, the papers that reported two different results with a sufficiently different dataset or methodology are \citet{Granett2009,Ferraro2015,Shajib2016,Cai2017,Kovacs2017}.

\begin{figure*}
\begin{center}
\includegraphics[width=\textwidth]{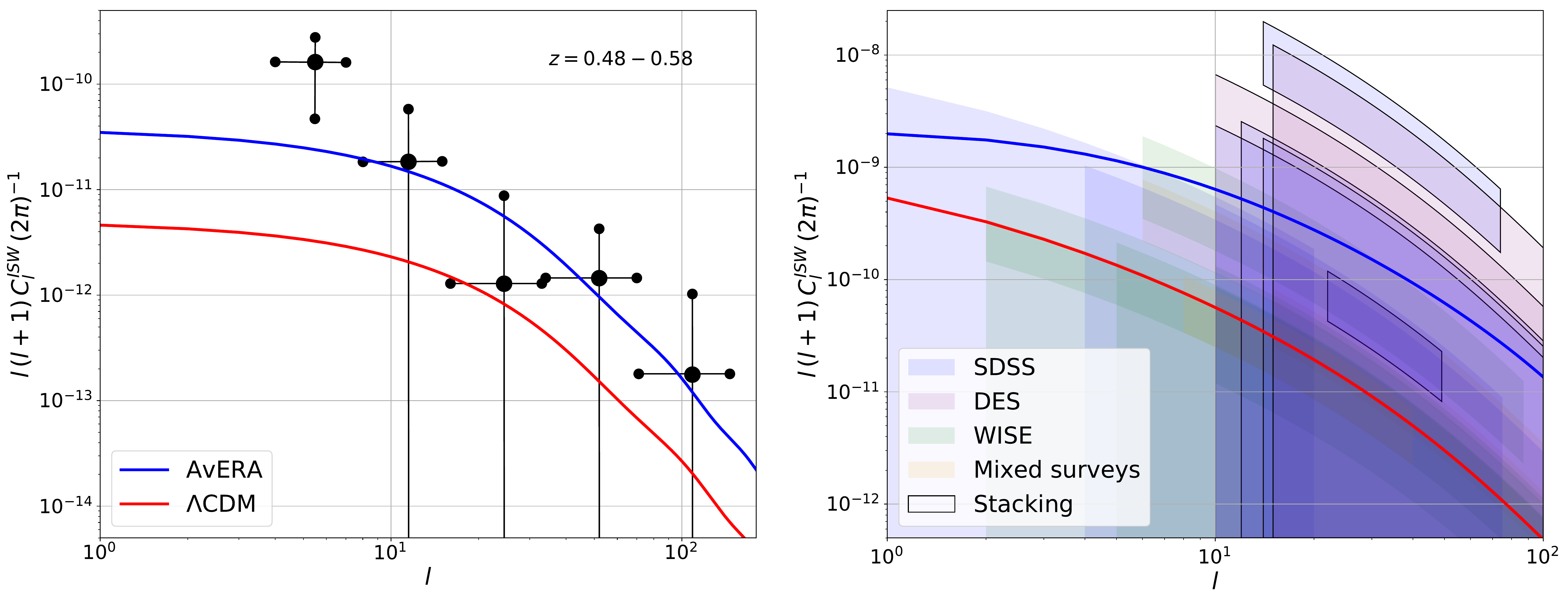}
\vspace*{-0.5cm}
\end{center}
\caption{ISW autocorrelation spherical power spectra for the AvERA (blue) and the standard $\Lambda \mathrm{CDM}$ cosmology (red). Left panel: power in the narrow redshift range $z=0.48-0.58$, alongside measured points and 1-sigma error bars from \citet{Granett2009}, in black. Right panel: various ISW measurements in the literature are indicated with shaded regions corresponding to 1-sigma ranges (see the text). The power was computed in the full redshift range of $z=0-8.55$. A black border around a given region denotes that the measurement was performed with the void--cluster stacking approach, while cross-correlation measurements lack a border. Note that due to the log-scale, the mean of the shaded areas is towards the top. See the text for a discussion of issues regarding the figure.}
\label{fig:ISWliterature}
\end{figure*}

Most of the works only compute either the CMB--matter cross-correlation for the given sample, or the stacked temperature signal in matched filters, but not the ISW autocorrelation as a function of $l$ that we presented in the previous section. To the best of our knowledge, \citet{Granett2009} is the only exception --- we extracted the measured data points from Fig.~7 of that paper, and computed the theoretical expectation for our two cosmologies in the corresponding narrow redshift range of $z=0.48-0.58$. Results are presented on the left panel of Fig.~\ref{fig:ISWliterature}. The AvERA cosmology lies closer to the points, but with such large error bars, both models are consistent with the data.

As for the rest of the measurements, the quantitative result that we \textit{can} compare to is the $A$ amplitude factor relative to $\Lambda \mathrm{CDM}$ expectations and its assumed Gaussian error, which corresponds to a best-fitting $\Lambda \mathrm{CDM}$ model scaled by $A$. However, we also require a range of $l$ values that the given measurement was sensitive to, since the AvERA--$\Lambda \mathrm{CDM}$ relationship was found to have strong dependence on $l$.

In the case of cross-correlation measurements, the determination of the $l$ range was relatively straightforward --- we extracted it from figures that showed points derived from data as a function of $l$, and/or used the reported cut-offs in $l$. When the $\alpha$ angular separation was presented instead of $l$, the simple approximation of $l=\frac{\pi}{\alpha}$ was used. For stacking measurements the process was more difficult: the angular power spectrum of the given filter shape was computed, and we chose the half-maximum range in $l$ as the effective range of the measurement. \citet{Granett2008} used a single, unscaled filter therefore this result was well-determined, but when the filters were scaled, we selected the $l$ range that covered the half-maximum ranges of filters both at the small and large edges of the filter radius distribution. In fact, we had to leave out the result of \citet{Nadathur2016} because the parameter distributions for the fitted filters were not provided, hence we could not determine an effective $l$ range.

\citet{Papai2011} had an extra void parameter, $R_v$, and we chose the $A$ 1-sigma range such that it covered the highest and lowest amplitude ratios observed within the 1-sigma range of $R_v$, also considering the $A$ 1-sigma ranges. Data points were extracted from the figures of the paper.

The right panel of Fig.~\ref{fig:ISWliterature} presents the AvERA and standard $\Lambda \mathrm{CDM}$ ISW power spectra along with observational results from the listed works. For each measurement, we show the $\Lambda \mathrm{CDM}$ curve scaled according to the given $A$ 1-sigma range, but only in the $l$ range the measurement was deemed sensitive to. Since we are dealing with autocorrelation spectra, the factor $A$ had to be squared, therefore the $\Lambda \mathrm{CDM}$ values were multiplied by the $68.27\%$ confidence intervals of $A^2$ (assuming that the distribution of $A$ is Gaussian). 

There are a number of caveats concerning the figure that have to be considered. First, measurements that utilize the same dataset (or indeed, datasets with overlapping spatial volume) cannot be regarded as independent. Second, due to the numerous approximations and simplifications mentioned above, the accuracy of the estimated $l$ coverage or even some $A$ values may be limited, and is heterogeneous across different measurements. Third, the weighting of different $l$ values in the measurements, i.e. which $l$-s dominate the fit in $A$, is entirely unclear.

Because of these severe limitations, we do not attempt to draw any quantitative conclusions. Instead, the right panel of Fig.~\ref{fig:ISWliterature} should be considered as a visual guide, highlighting the lack of consensus among ISW measurements in the literature. While most stacking approaches appear to favor the AvERA model, those have very large error bars, and numerous cross-correlation results are bundled around the $\Lambda \mathrm{CDM}$ curve, with smaller errors. However, the question of which cosmological model fits the data better cannot be quantitatively evaluated without reproducing the analyses entirely.

\section{Discussion}

In the current paper, we investigated the magnitude of the ISW effect predicted by the AvERA algorithm, relative to the standard $\Lambda \mathrm{CDM}$ cosmological model. The former claimed to solve the apparent tension between local and CMB-based measurements of the Hubble constant \citep{Riess2016}, and produced a similar expansion history to that of $\Lambda \mathrm{CDM}$ without assuming dark energy \citep{Racz2017}. The ISW effect is an excellent probe of the finer details of expansion histories, therefore it should be a convenient tool to perform a detailed test of the merits of cosmologies, to support or even disprove theories. However, observationally the field is still evolving, with limited clarity in results.

Using both a raytracing and theoretical approach, we have concluded that, in the context of the linear growth approximation, the AvERA algorithm predicts an ISW effect significantly larger than the $\Lambda \mathrm{CDM}$ expectation, with a relative amplitude factor between $A=1.93-5.29$, depending on the $l$ index of the spherical power spectrum (Sect.~\ref{sec:amplitude} and Fig.~\ref{fig:ISWpower}). \citet{Afshordi2004} showed that the theoretical maximum significance detection for the ISW effect is $\sim 7.5 \sigma$, while an all-sky survey with $10$ million galaxies within $0 < z < 1$ would yield a $\sim 5.0 \sigma$ detection. Taking these numbers at face value, an ISW effect with twice the amplitude --- the lower bound of the factor we found for AvERA, at low $l$-s --- would be differentiable from the $\Lambda \mathrm{CDM}$ prediction at a similarly high level of significance.

We also investigated how the predicted relative amplitude factor scales with the particle mass input parameter of AvERA, and found only a slight dependence, not significant to impact any of the conclusions (refer to Appendix~\ref{sec:appendix2} for more details). 

Furthermore, we found that the AvERA cosmology suggests a smaller, opposite-sign ISW effect in the redshift range $z \approx 1.5 - 4.4$, the presence of which could be a \textit{smoking gun}, signaling a departure from $\Lambda \mathrm{CDM}$, or alternatively, the absence of which could disprove the AvERA model (Sect.~\ref{sec:cosmo}). A high-redshift quasar sample with sufficiently large sky coverage and sample size could potentially yield a detection of such an effect.

Additionally, we contrasted our results with various previous observational measurements of the ISW effect. Our comparison highlighted the significant scatter and slight to moderate tensions between different works (Sect.~\ref{sec:literature} and Fig.~\ref{fig:ISWliterature}). We refrained from attempting a quantitative analysis due to a number of limitations and difficulties in aggregating heterogeneous measurements. We avoided drawing even qualitative conclusions, as they would depend on subjectively ,,cherry-picking'' from among the observations.

In any case, once the observational results have been consolidated and consensus has been reached, the power spectrum of the ISW effect will be a decisive test of the AvERA and $\Lambda \mathrm{CDM}$ cosmological models, as their predictions differ significantly. Newer surveys that are deeper and wider than current counterparts, e.g. Pan-STARRS \citep{Tonry2012}, LSST \citep{Ivezic2008} or Euclid \mbox{\citep{Amiaux2012}} should yield data suitable for such analyses. Alternatively, a complete and careful reanalysis of existing data might also deliver definitive answers.

\section{Acknowledgements}

IS and RB acknowledge support from the National Science Foundation (NSF) award 1616974. IC, GR and RB were supported by the NKFI NN 114560 grant of Hungary.

\bibliographystyle{mn2e}
\bibliography{avera_isw}

\appendix

\section{On the accuracy of the Limber approximation}

\label{sec:appendix1}

The Limber approximation is frequently used in the literature to simplify the calculation of integrals in spherical projections \citep{Limber1953,Kaiser1992,Ho2008,Cai2009}. It assumes small angular separations (i.e. large $l$ values), and also assumes that some inner factors in the integrals are only slowly changing. We can use the approximation to simplify Eqs.~\ref{eq:sphericalpower}-\ref{eq:bessel} by setting
\begin{equation}
k = \frac{l+1/2}{\chi},
\label{eq:limber1}
\end{equation}
and writing
\begin{equation}
\frac{2}{\pi} \int dk \, k^2 j_l[\chi k] \, j_l[\chi' k] \approx \frac{1}{\chi^2} \delta(\chi-\chi').
\label{eq:limber2}
\end{equation}
Substituting these into Eqs.~\ref{eq:sphericalpower}-\ref{eq:bessel}, we get the Limber approximation formula for the theoretical ISW calculation,
\begin{multline}
C_l^{\mathrm{ISW}} = T_{\mathrm{CMB}}^2 \left(\frac{3 H_0^2\Omega_m}{c^2}\right)^2 \frac{1}{\left(l+1/2\right)^4} \times \\ \times \int d \tau \, \frac{\chi^2(\tau)}{c} \left[ \frac{d}{d\tau}\left( \frac{D}{a} \right) \right]^2 P_\delta \left(\frac{l+1/2}{\chi(\tau)}\right).
\label{eq:limber3}
\end{multline}

\begin{figure}
\begin{center}
\includegraphics[width=\columnwidth]{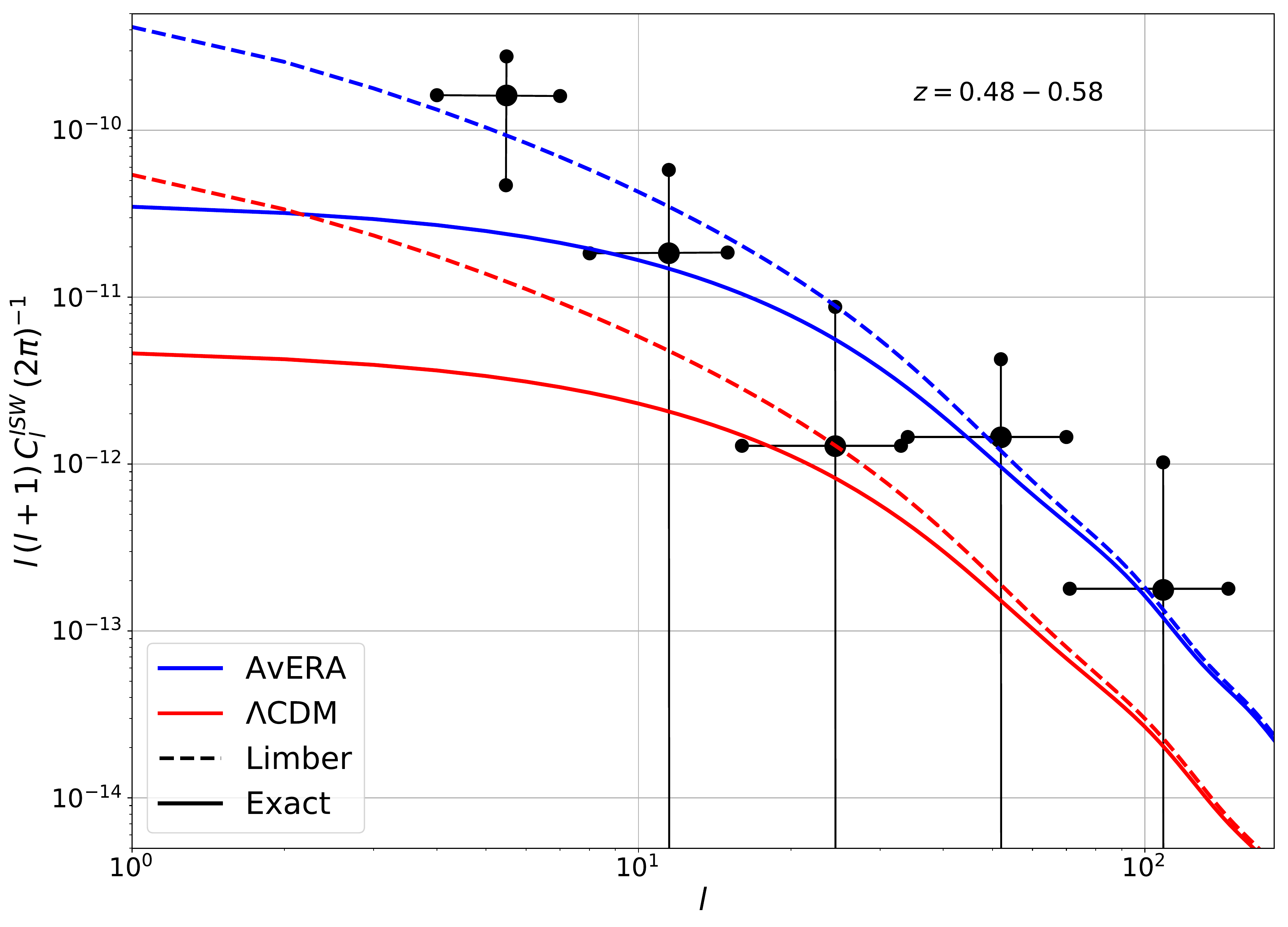}
\vspace*{-0.5cm}
\end{center}
\caption{The same as the left panel of Fig.~\ref{fig:ISWliterature}, with solid lines indicating the results of the full theoretical ISW calculation, and the additional dashed lines representing the Limber approximation, for the narrow redshift range $z=0.48-0.58$.}
\label{fig:Limber1}
\end{figure}

\citet{Loverde2008} showed that the Limber approximation has limited accuracy when applied to galaxy--galaxy cross-correlation in a narrow redshift bin. We have verified this result to also hold for the ISW autocorrelation power spectrum, illustrated by Fig.~\ref{fig:Limber1} --- in the redshift range $z=0.48-0.58$, corresponding to the coverage of the sample of \citet{Granett2009}, the Limber curve starts to diverge from the full expression at $l\approx 40$, for both the AvERA and $\Lambda \mathrm{CDM}$ cosmologies.

\begin{figure}
\begin{center}
\includegraphics[width=\columnwidth]{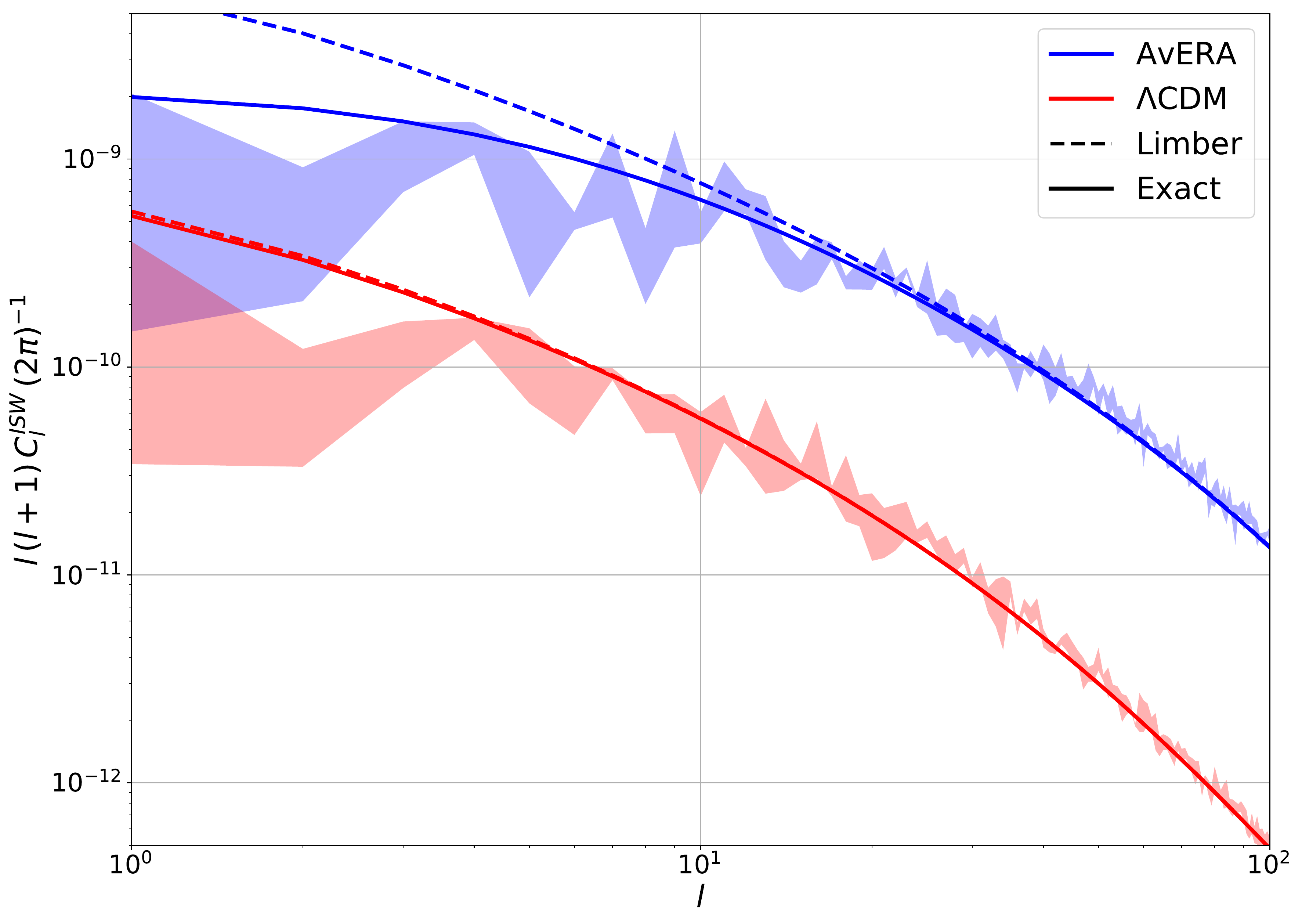}
\vspace*{-0.5cm}
\end{center}
\caption{The same as Fig.~\ref{fig:ISWpower}, with solid lines indicating the results of the full theoretical ISW calculation, and the additional dashed lines representing the Limber approximation, for the entire redshift range of $z=0-8.55$.}
\label{fig:Limber2}
\end{figure}

However, there is another issue concerning the Limber approximation that is unique to the AvERA cosmology. From Eq.~\ref{eq:limber3} it is clear that the factor $\frac{d}{d\tau}\left( \frac{D}{a} \right)$ became squared due to applying the approximation. Additionally, Fig.~\ref{fig:cosmo} shows that this factor has an opposite-sign component at a higher redshift range for AvERA. Thus, in the Limber approximation, this component raises the total ISW power due to the square, while the full calculation correctly reflects that the total ISW signal is reduced by the opposite-sign contribution. Fig.~\ref{fig:Limber2} shows that in the full $z=0-8.55$ redshift range, the Limber curve visibly differs from the complete result below $l\approx 10$ in the case of AvERA, while being significantly more accurate for the $\Lambda \mathrm{CDM}$ model.

We conclude that care should be taken when applying the Limber approximation in situations where its accuracy has not yet been thoroughly evaluated.

\section{Dependence on the particle mass parameter}

\label{sec:appendix2}

In Sect.~\ref{sec:cosmo}, we discussed that the particle mass input parameter of the AvERA simulation affected the local value of the Hubble constant.

We investigated whether there is a similar dependence for the ISW autocorrelation power spectrum by evaluating Eqs.~\ref{eq:sphericalpower}-\ref{eq:bessel}, using AvERA cosmologies corresponding to the most likely mass parameter values given the $H_0$ measurements of \citet{Riess2016}. The particle mass of $1.17 \times 10^{11} M_\odot$ yielded an $H_0$ that approximately matched the measurement, while the values $2.03 \times 10^{11} M_\odot$ and $3.96 \times 10^{11} M_\odot$ corresponded to an $H_0$ close to 1 and 2 sigmas below that of \citet{Riess2016}, respectively \citep[refer to Fig.~4 and Table 1 in][]{Racz2017}.

\begin{figure}
\begin{center}
\includegraphics[width=\columnwidth]{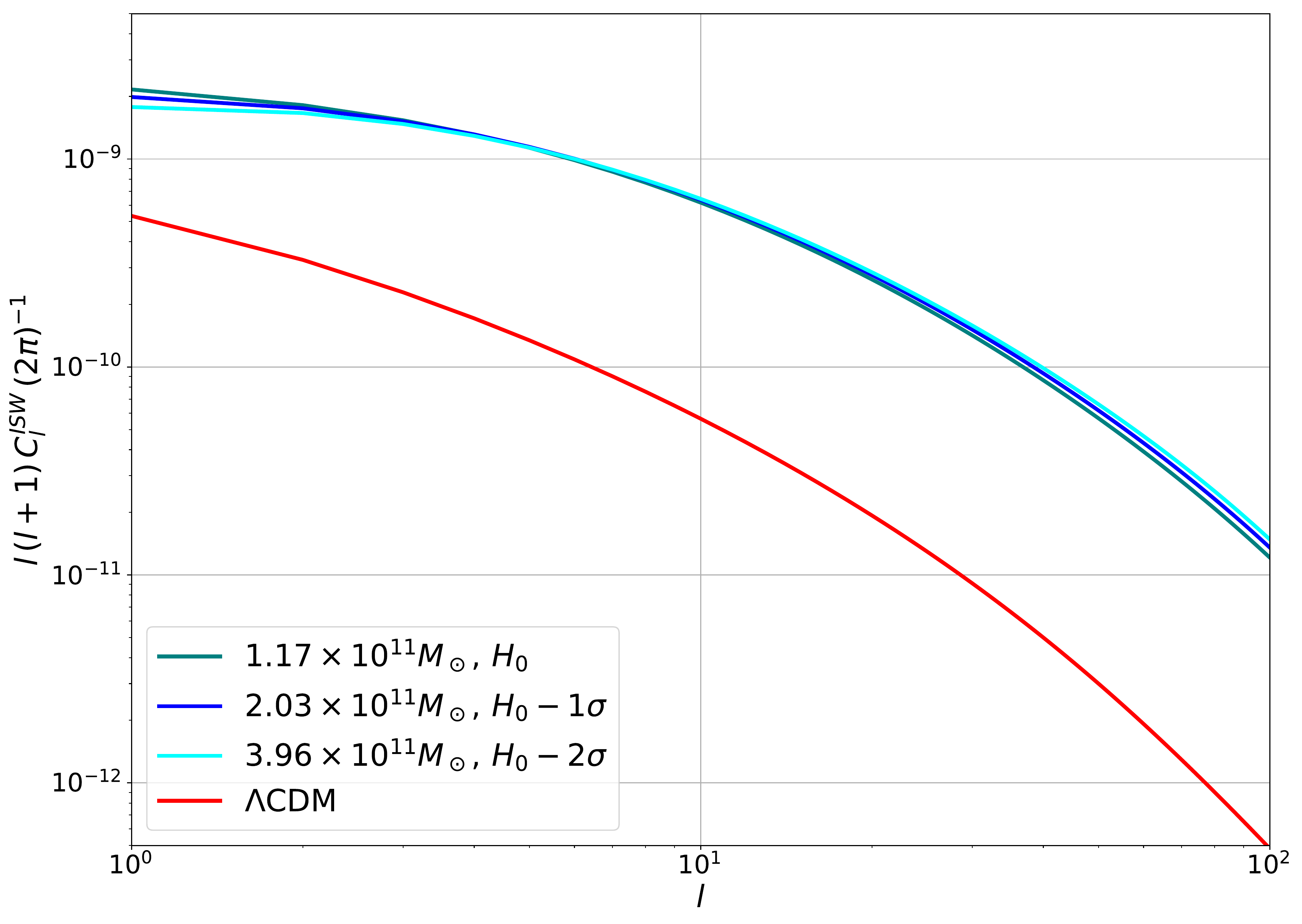}
\vspace*{-0.5cm}
\end{center}
\caption{The same as Fig.~\ref{fig:ISWpower}, but with the raytracing results omitted. Different shades of blue indicate AvERA power spectra for different particle mass input parameters --- the mass where the $H_0$ measurement of \citet{Riess2016} is approximately matched, 1 sigma below that value, and finally 2 sigmas below the local $H_0$.}
\label{fig:particlemass}
\end{figure}

The results are shown in Fig.~\ref{fig:particlemass}. It is clear that the predicted ISW signal does not depend strongly on the particle mass, the amplitude factor range of $A=1.83-5.54$ (relative to $\Lambda \mathrm{CDM}$) covers all observed values, which should be contrasted with our original range $A=1.93-5.29$. Thus, our conclusions are not affected by the choice of this parameter.

\end{document}